\documentclass[aps]{revtex4-1}
\usepackage{amssymb}
\usepackage{amsmath}
\usepackage{bm}
\usepackage[english]{babel}
\usepackage{subfigure}
\usepackage{graphicx}
\usepackage{color}

\begin{document}

\title{The transition from the classical to the quantum regime in nonlinear
Landau damping. }
\author{$^1$G. Brodin, $^1$J. Zamanian, $^2$J. T. Mendonca}
\affiliation{$^1$Department of Physics, Ume{\aa } University, SE--901 87
Ume{\aa}, Sweden, \\ $^2$IPFN, Instituto Superior Tecnico, 1049-001 Lisboa, Portugal}

\begin{abstract}
Starting from the Wigner-Moyal equation coupled to Poisson's equation, a
simplified set of equations describing nonlinear Landau damping of Langmuir
waves is derived. This system is studied numerically, with a particular
focus on the transition from the classical to the quantum regime. In the
quantum regime several new features are found. This includes a quantum
modified bounce frequency, and the discovery that bounce-like amplitude
oscillations can take place even in the absence of trapped particles. The
implications of our results are discussed.
\end{abstract}
\keywords{Quantum plasma, Wave-particle interaction, Bounce oscillations}
\maketitle

\section{Introduction}

Numerous aspects of quantum plasmas have been investigated during the last
decade, see e.g. \cite{manfredi2006,Haas-book,Shukla-Eliasson-RMP}. Basic
features such as electron degeneracy and particle dispersive properties have
been studied in some detail \cite%
{manfredi2006,Haas-book,Shukla-Eliasson-RMP,Bret-2007,Andreev-2011,Vladimirov-2013,Brodin-2013,Surface-q,Hall-MHD-q}
Exchange effects \cite{Manfredi-DFT,Zamanian-exchange}, the magnetic dipole
force and other contributions from the electron spin \cite%
{Zamanian-2010-NJP,Lundin-2010} have also been examined, including
relativistic effects \cite{Asenjo-2011,Mend-2011}. Systems of interest in
this context include e.g. quantum wells \cite{Manfredi-quantum-well},
laser-plama interaction on solid density targets \cite{glenzer-redmer}, and
astrophysical plasmas \cite{Astro}. Works on quantum plasmas also have
relevance for recent applications in plasmonics \cite{Atwater-Plasmonics}
and spintronics \cite{Spintronics}.

In the present paper we will consider the influence of quantum effects on
the nonlinear regime of Landau damping, based on the Wigner-Moyal equation 
\cite{Haas-book} that accounts for particle dispersive effects. Previous
works in this area \cite{SFR-1991,Daligault-2014} have deduced that quantum
effects can suppress the nonlinear bounce oscillations of Langmuir waves and
turn the evolution into basic linear damping. Moreover, quantum corrections
to electron holes in phase space that may form as a result of wave-particle
interaction have been calculated in Ref. \cite{Fedele-2004}, and the
quasilinear theory of the Wigner-Poisson system has been studied in Ref. 
\cite{Haas-2008}. However, many of the details of nonlinear Landau damping
have not been studied before in the quantum regime. Making analytical
approximations applicable for a resonance in the tail of the distribution,
we first simplify the Wigner-Moyal equation coupled to Poisson's equation
into a system that is more easy to solve numerically. This system is shown
to fulfill an energy conservation law, and reduces to a previously studied
system \cite{Brodin-1997} in the classical limit. A systematic study of the
transition from classical to quantum behavior then reveals several new
features. This include a quantum modification of the bounce frequency, a new
condition for the quantum suppression of the nonlinear regime and the
discovery that bounce-like oscillations can take place even in the absence
of trapped particles.

Similar to most cases, the conditions needed for quantum effects to be
important in our problem include a high plasma density and a modest plasma
temperature. However, while the scaling with temperature and density is the
same as usual \cite{manfredi2006,Shukla-Eliasson-RMP}, the precise numerical
values are to some extent relaxed compared to the standard expressions in
case the resonance lies in the tail of the distribution. The general
conclusion is that the properties of wave-particle interaction are more
easily influenced by quantum effects than the ordinary fluid properties. A
concrete example is provided in the final section, and the significance of
our results are discussed.

\section{Basic equations and derivations}

Our starting point is the Wigner-Moyal equation \cite{Haas-book}, which reads

\begin{equation}
\frac{\partial f}{\partial t}+\mathbf{v}\cdot \nabla f-\frac{iq}{\hbar }\int 
\frac{d^{3}\mathbf{r}^{\prime }d^{3}\mathbf{v}^{\prime }m^{3}}{(2\pi \hbar
)^{3}}e^{i\mathbf{r}^{\prime }\cdot (\mathbf{v-v}^{\prime })m/\hbar }\left[
\Phi (\mathbf{r}+\mathbf{r}^{\prime }/2)-\Phi (\mathbf{r}-\mathbf{r}^{\prime
}/2)\right] f(\mathbf{r,v}^{\prime },t)=0  \label{Wigner-1}
\end{equation}%
Here $f$ is the Wigner-function, $\Phi $ is the electrostatic potential, $q$
(=$-e$) is the electron charge, $m$ is the electron mass and $h=2\pi \hbar $
is Planck's constant. Eq. (\ref{Wigner-1}) is combined with Poisson's
equation 
\begin{equation}
-\nabla ^{2}\Phi =\frac{q}{\varepsilon _{0}}\int fd^{3}\mathbf{v}
\label{Poisson}
\end{equation}%
to give us a closed set. Eq. (\ref{Wigner-1}) applies for electrostatic
fields and do not account for the spin of the electrons or exchange effects.
For generalizations to electromagnetic fields and spin effects, see e.g.
Ref. \cite{Zamanian-2010-NJP} and for generalizations to include exchange
effects, see e.g. \cite{Zamanian-exchange}. Here we focus on the problem of
Langmuir waves in an unmagnetized plasma, in which case the omissions of
electromagnetic effects and spin effects are trivially justified.
Furthermore, we will consider the case of a moderate plasma density with $%
\hbar \omega _{p}\ll k_{B}T$. It is then safe to exclude exchange effects 
\cite{Zamanian-exchange}. In fact, the condition $\hbar \omega _{p}\ll
k_{B}T $ is often used also to disregard the particle dispersive quantum
effects included in Eq. (\ref{Wigner-1}). Such a condition is indeed an
appropriate one for neglecting e.g. the particle dispersive contribution to
the real part of the Langmuir dispersion relation and for neglecting the
quantum contribution of Eq. (\ref{Wigner-1}) in many other cases \cite%
{manfredi2006,Haas-book,Shukla-Eliasson-RMP}. However, as we will see below,
the quantum contribution to wave particle interaction can be crucial even in
the regime $\hbar \omega _{p}\ll k_{B}T$. \ A useful starting point to see
this is to rewrite (\ref{Wigner-1}) in the alternative way \cite%
{Mendonca-2001} 
\begin{equation}
\frac{\partial f}{\partial t}+\mathbf{v}\cdot \nabla -\frac{q}{m}\Phi \cdot %
\left[ \frac{2m}{\hbar }\sin \left( \frac{\hbar }{2m}\frac{\overleftarrow{%
\partial }}{\partial \mathbf{r}}\cdot \frac{\overrightarrow{\partial }}{%
\partial \mathbf{v}}\right) f\right] =0.  \label{Wigner-2}
\end{equation}%
Here the arrows on the operators indicate in which direction they are
acting, and the sinus-operator is defined in terms of its Taylor-expansion.
Let us first consider linear plane wave solutions $\propto \exp i(kz-\omega
t)$, and estimate the relative importance of quantum corrections. Letting $%
f=f_{0}+\hat{f}_{1}\exp i(kz-\omega t)$ we can use the classical linear
solution 
\begin{equation}
\hat{f}_{1}=\frac{qk\hat{\Phi}}{m(\omega -k_{z}v_{z})}\frac{\partial f_{0}}{%
\partial v_{z}}
\end{equation}%
as a means to illustrative the importance of quantum corrections. Comparing
the magnitude of the first order quantum correction in the Taylor expansion
with that of the classical term, we note that in the bulk of the velocity
distribution (i.e. for $v\sim v_{t}$) the quantum corrections are small
provided $Q_{\mathrm{bulk}}\ll 1$, where 
\begin{equation}
Q_{\mathrm{bulk}}=\frac{\hbar k}{mv_{t}}.
\end{equation}%
However, looking at velocities close to the wave-particle resonance we must
let $\omega -k_{z}v_{z}$ be small, of the order $\ \omega -k_{z}v_{z}\sim
\gamma _{L}$ where $\gamma _{L}$ is the linear damping rate. Close to the
resonance the relative importance of the quantum terms are given by the
parameter $Q_{\mathrm{res}}$, which is 
\begin{equation}
Q_{\mathrm{res}}=\frac{\hbar k^{2}}{m\gamma _{L}}.
\end{equation}%
For a resonance in the tail of the distribution $\gamma _{L}/k\ll v_{t}$ in
which case we may have $Q_{\mathrm{res}}\sim 1$ at the same time as $Q_{%
\mathrm{bulk}}$ $\ll 1$. This regime will be the focus in what follows. As a
result it suffices to solve the Vlasov limit of Eq. (\ref{Wigner-2}) in most
of velocity space, but close to the resonance we need to solve the full
Wigner equation. A similar approach applies for the small amplitude
approximations. Provided $qk\hat{\Phi}/m\omega v_{t}\ll 1$ we have $%
\left\vert \partial \hat{f}_{1}/\partial v_{z}\right\vert \ll \left\vert
\partial f_{0}/\partial v_{z}\right\vert $ unless we are close to the
resonance. We will assume $qk\hat{\Phi}/m\omega v_{t}\ll 1$ to hold, and
thus for most of velocity space we can solve the \textit{linearized Vlasov
relation}. However, close to the resonance where we can have nonlinear
wave-particle interaction we must solve the \textit{Wigner equation without
making linear approximations}.

For a resonance in the tail of the distribution the number of particles
contributing to the nonlinear interaction are relatively few. As a result
there will be only minor harmonic generation of the electric field.
Consequently we will use the ansatz of a slowly time-varying plane wave $%
\Phi =\hat{\Phi}(t)\exp i(kz-\omega t)+\mathrm{c.c}.$ for the potential,
where c.c. stands for complex conjugate In general the Wigner function will
be a periodic function, that will be represented as 
\begin{equation*}
f=f_{0}(\mathbf{v})+\delta f_{0}(\mathbf{v,}t)+\left[ \sum_{n=1}^{\infty
}f_{n}(\mathbf{v,}t)\exp i[n(kz-\omega t)]+\mathrm{c.c}\right] .
\end{equation*}%
Next we divide the velocity space into the resonant region $[v_{z}-\delta v_{%
\mathrm{res}},v_{z}+\delta v_{\mathrm{res}}]$ and the non-resonant region
containing the complementary part. \ When evaluating the charge density in (%
\ref{Poisson}) we thus use $\int fd^{3}v=\int_{\mathrm{nr}}fd^{3}v+\int_{%
\mathrm{res}}fd^{3}v$ where the subscripts nr and res denotes the
nonresonant and resonant regions, respectively. Since the linear Vlasov
equation applies in the nonresonant region we can use%
\begin{eqnarray}
\int_{\mathrm{nr}}fd^{3}v &=&\int_{\mathrm{nr}}f_{1}(\mathbf{v}%
,t)e^{i(kz-\omega t)}d^{3}v+\mathrm{c.c}=  \notag \\
&=&\frac{qk\hat{\Phi}(t)}{m}\int_{\mathrm{nr}}\frac{e^{i(kz-\omega
t)}(\partial f_{0}/\partial v_{z})}{(\omega -k_{z}v_{z})}d^{3}v-\int_{%
\mathrm{nr}}\frac{e^{i(kz-\omega t)}(\partial f_{1}/\partial t)}{i(\omega
-k_{z}v_{z})}d^{3}v+\mathrm{c.c.}  \label{Non-res-approx}
\end{eqnarray}%
Noting that the second term on the right hand side of (\ref{Non-res-approx})
is a small correction, proportional to the slowly varying amplitude, we can
use the lowest order approximation (i.e. dropping time derivatives on the
amplitude) to convert it to a term proportional to $\partial \hat{\Phi}%
(t)/\partial t$.\ Combining (\ref{Non-res-approx}) with (\ref{Poisson}) we
then deduce 
\begin{equation}
\frac{\partial \hat{\Phi}(t)}{\partial t}=\frac{q}{\varepsilon
_{0}k^{2}\partial D/\partial \omega }\int_{\mathrm{res}}f_{1}d^{3}v
\label{Phi-evolution-1}
\end{equation}%
where $D(\omega ,k)=1+(q^{2}/km\varepsilon _{0})$ $\int_{\mathrm{nr}%
}(\partial f_{0}/\partial v_{z})d^{3}v/(\omega -k_{z}v_{z})$. Note that the
dispersion function $D(\omega ,k)$ has a component of arbitrariness in the
definition, since it depend on the width of the resonant region $\delta v_{%
\mathrm{res}}$. Nevertheless in the derivation of (\ref{Phi-evolution-1}) we
have taken $D(\omega ,k)=0$ to hold by definition (i.e. $\omega (k)$ is
defined by this relation), and in case a possible frequency shift occurs due
to this, it is included in the time-dependence of $\hat{\Phi}\,$. It should
be stressed that although a large number, of harmonics $f_{n}(\mathbf{v,}t)$
might be needed to solve for the Wigner function in the resonant region, it
is only the first harmonic $f_{1}$ that contributes in (\ref{Phi-evolution-1}%
).

Next we need to solve for the Wigner function in the resonant region. We
restrict ourselves to a Maxwellian background distribution $F_{0}(\mathbf{v}%
) $, and introduce the 1D-Wigner function 
\begin{equation}
g=G_{0}(v_{z})+g_{0}(v_{z}\mathbf{,}t)+\left[ \sum_{n=1}^{\infty }g_{n}(v_{z}%
\mathbf{,}t)\exp i[n(kz-\omega t)]+\mathrm{c.c}\right]
\label{Harmonic ansatz}
\end{equation}%
where $F_{0}(\mathbf{v})=G_{0}(v_{z})\exp [(-v_{x}^{2}-v_{y}^{2})/v_{t}^{2}]$
and $f_{n}(\mathbf{v},t)=g_{n}(v_{z},t)\exp
[(-v_{x}^{2}-v_{y}^{2})/v_{t}^{2}]$. This ansatz is then substituted into (%
\ref{Wigner-2}). Since the gradient operator becomes $\pm ik\mathbf{\hat{z}}$
(as the spatial dependence of the potential is $\exp (\pm ikz)$) a useful
formula is 
\begin{equation}
\frac{2m}{\hbar }\sin \left( \pm \frac{i\hbar k}{2m}\frac{\partial }{%
\partial v_{z}}\right) g_{n}(v_{z},t)=\frac{\pm im\left[ g_{n}(v_{z}+\hbar
k/2m)-g_{n}(v_{z}-h\hbar /2m)\right] }{\hbar }.  \label{Sinus-operator}
\end{equation}%
With the help of this formula, and subsituting the ansatz (\ref{Harmonic
ansatz}) into (\ref{Wigner-2}), the following set of coupled equations are
deduced%
\begin{eqnarray}
\frac{\partial g_{1}}{\partial t}-i\delta \omega (v_{z})g_{1} &=&\frac{q}{%
\hbar }\hat{\Phi}[G_{0}(v_{z}+\hbar k/2m)-G_{0}(v_{z}-\hbar
k/2m)+g_{0}(v_{z}+\hbar k/2m)-g_{0}(v_{z}-\hbar k/2m)]  \notag \\
&&+\hat{\Phi}^{\ast }[g_{2}(v_{z}+\hbar k/2m)-g_{2}(v_{z}-\hbar k/2m)]
\label{g1new} \\
\frac{\partial g_{n}}{\partial t}-in\delta \omega (v_{z})g_{n} &=&\frac{q}{%
\hbar }\hat{\Phi}[g_{n-1}^{\ast }(v_{z}+\hbar k/2m)-g_{n-1}^{\ast
}(v_{z}-\hbar k/2m)]+\hat{\Phi}^{\ast }[g_{n+1}(v_{z}+\hbar
k/2m)-g_{n+1}(v_{z}-\hbar k/2m)]  \label{gn}
\end{eqnarray}%
and

\begin{equation}
\frac{\partial g_{0}}{\partial t}=\frac{q}{\hbar }\hat{\Phi}[g_{1}^{\ast
}(v_{z}+\hbar k/2m)-g_{1}^{\ast }(v_{z}-\hbar k/2m)]+\hat{\Phi}^{\ast
}[g_{1}(v_{z}+\hbar k/2m)-g_{1}(v_{z}-\hbar k/2m)]  \label{g0}
\end{equation}%
where $\delta \omega (v_{z})=\omega -kv_{z}$ and the star denotes complex
conjugate. Using $\int_{\mathrm{res}}f_{1}d^{3}v=\pi v_{t}^{2}\int_{\mathrm{%
res}}g_{1}dv_{z}$ in (\ref{Phi-evolution-1}) we see that Eqs. (\ref{g1new})-(%
\ref{g0}) and (\ref{Phi-evolution-1}) constitute a closed set. The equations
agree with those studied by Ref. \cite{Brodin-1997} in the classical limit $%
\hbar \rightarrow 0$, provided the collisional frequency is put to zero in
that work. It is clear that the quantum features are encoded in the velocity
shift $\hbar k/2m$. \ An important thing to note is that we can use $%
G_{0}(v_{z}+\hbar k/2m)-G_{0}(v_{z}-\hbar k/2m)\approx \lbrack \partial
G_{0}(v_{z})/\partial v_{z}]\hbar k/m$ for $\hbar \omega /mv_{t}^{2}\ll 1$,
whereas similar approximations are not applicable for the perturbed Wigner
function, as the quantities $g_{n}$ varies on a much shorter scale length.

Before we can proceed we must put constraints on the parameter $\delta v_{%
\mathrm{res}}$. For the resonant region to cover a sufficient amount of
resonant particles in the linear regime we need the condition $\delta v_{%
\mathrm{res}}\gg \gamma _{L}/k$. Moreover, to cover particles that are
closed to be trapped in the potential well with a sufficiently large margin
we need $\delta v_{\mathrm{res}}\gg \omega _{B}/k$, where $\omega
_{B}=(qk^{2}\Phi /m)^{1/2}$ is the bounce frequency of trapped particles
(with this choice the the resonant region is at least an order of magnitude
larger than the region of trapped particles). Finally, to cover the quantum
effects properly we need $\delta v_{\mathrm{res}}\gg \hbar k/2m$. At the
same time the calculation scheme is based on the resonance region being much
smaller than the thermal velocity, and hence we need $\delta v_{\mathrm{res}%
}\ll v_{t}$. If we sharpen this condition slightly and limit ourselves to $%
\delta v_{\mathrm{res}}\ll kv_{t}^{2}/\omega $ we may take $[\partial
G_{0}(v_{z})/\partial v_{z}]$ as constant in the resonance region ($%
=[\partial G_{0}(v_{z})/\partial v_{z}]_{\omega /k}$), which simplify some
of the technical aspects. For a resonance that lies in the tail of the
distribution we can have $kv_{t}/\gamma _{L}\gtrapprox 100$, in which case
it is easy to fulfill all conditions simultaneously, unless the nonlinearity
or the quantum effects are extremely strong. Importantly, the numerical
solutions presented below are not dependent on the precise choice of $\delta
v_{\mathrm{res}}$, as long as the above conditions are fulfilled.

Next we introduce normalized variables. Choosing normalized time as $\gamma
_{L}t$, normalized velocity as $kv_{z}/\gamma _{L}$, normalized potential as 
$qk^{2}\Phi /\gamma _{L}^{2}m$ and normalized harmonics of the Wigner
function as 
\begin{equation*}
\frac{kg_{n}}{\gamma _{L}[\partial G_{0}(v_{z})/\partial v_{z}]_{\omega /k}}
\end{equation*}%
the coupled equations become 
\begin{equation}
\frac{\partial \hat{\Phi}(t)}{\partial t}=\frac{1}{\pi }\int_{\mathrm{res}%
}g_{1}dv  \label{Norm-1}
\end{equation}%
\begin{eqnarray}
\frac{\partial g_{1}}{\partial t}+ivg_{1} &=&\hat{\Phi}\left[ 1+\frac{%
g_{0}(v+\delta v_{q})-g_{0}(v-\delta v_{q})}{2\delta v_{q}}\right] +\hat{\Phi%
}^{\ast }\left[ \frac{g_{2}(v+\delta v_{q})-g_{2}(v-\delta v_{q})}{2\delta
v_{q}}\right]  \label{Norm-2} \\
\frac{\partial g_{n}}{\partial t}+invg_{n} &=&\hat{\Phi}\left[ \frac{%
g_{n-1}^{\ast }(v+\delta v_{q})-g_{n-1}^{\ast }(v-\delta v_{q})}{2\delta
v_{q}}\right] +\hat{\Phi}^{\ast }\left[ \frac{g_{n+1}(v+\delta
v_{q})-g_{n+1}(v-\delta v_{q})}{2\delta v_{q}}\right]  \label{Norm-3}
\end{eqnarray}%
and

\begin{equation}
\frac{\partial g_{0}}{\partial t}=\hat{\Phi}\left[ \frac{g_{1}^{\ast
}(v+\delta v_{q})-g_{1}^{\ast }(v-\delta v_{q})}{2\delta v_{q}}\right] +\hat{%
\Phi}^{\ast }\left[ \frac{g_{1}(v+\delta v_{q})-g_{1}(v-\delta v_{q})}{%
2\delta v_{q}}\right]  \label{Norm-4}
\end{equation}%
where $n\geq 2$ in Eq. (\ref{Norm-3}) and the quantum velocity shift is $%
\delta v_{q}=hk^{2}/2m\gamma _{L}$. We have omitted indices denoting
normalized variables for notational convenience. Provided the conditions for
the resonance region presented above is fulfilled, we may neglect
perturbations at the boundary, i.e. use the approximation $g_{n}(\pm \delta
v_{\mathrm{res}})\approx 0$ where $n=0,1,2...$.This property holds in the
vicinity of the boundary (where the vicinity means a velocity of the order $%
\delta v_{q}$), which means that 
\begin{equation}
\int_{\mathrm{res}}g_{0}(v+\delta v_{q})g_{1}^{\ast }(v)dv\approx \int_{%
\mathrm{res}}g_{0}(v)g_{1}^{\ast }(v-dv_{q})dv.  \label{boundary}
\end{equation}%
When $\delta v_{\mathrm{res}}$ is chosen large enough such that Eq. (\ref%
{boundary}) is fulfilled, as well as similar types of approximations
involving $g_{n}$, the system (\ref{Norm-1})-(\ref{Norm-4}) posses an energy
conservation law

\begin{equation}
\frac{\partial W_{\mathrm{tot}}}{\partial t}=\frac{\partial }{\partial t}%
\left[ \left\vert \hat{\Phi}(t)\right\vert ^{2}+\frac{1}{2}\int_{\mathrm{res}%
}g_{0}^{2}dv+\sum_{n=1}^{\infty }\int_{\mathrm{res}}\left\vert
g_{n}\right\vert ^{2}dv\right] =0  \label{energy}
\end{equation}%
where $W_{\mathrm{tot}}$ is the total energy. The first term of Eq. (\ref%
{energy}) represents the wave energy (including the kinetic energy in the
non-resonant region), whereas $\int_{\mathrm{res}}\left\vert
g_{n}\right\vert ^{2}dv$ represents the particle energy in the resonant
region of each harmonic. Note that the zero:th harmonic get an extra factor $%
1/2$, which is related to $g_{0}$ being real. In the numerical calculation
made in the next section Eq. (\ref{energy}) has been used as test of the
numerical scheme and a confirmation that $\delta v_{\mathrm{res}}$ has been
chosen large enough. In particular it should be stressed that the quantities 
$\int_{\mathrm{res}}\left\vert g_{n}\right\vert ^{2}dv$ is not sensitive to
the exact choice of $\delta v_{\mathrm{res}}$, since the integrand falls of
rapidly away from the resonance. This can tested by varying the parameter $%
\delta v_{\mathrm{res}}$ in the numerical code. We find that the relative
values of the wave energy and resonant energy changes around $0.001-0.002$
when the value of $\delta v_{\mathrm{res}}$ is changed a factor $1-3$ within
the bounds of the strong inequalities.

An advantage with the above system (\ref{Norm-1})-(\ref{Norm-4}) is that
solving the equations numerically we can follow the evolution taking
time-steps that are larger than the inverse plasma frequency, as the
equations contain only the slow time-scales, as opposed to the original
Wigner-Moyal equation (\ref{Wigner-2}). Moreover, we only need to solve the
equations in a small part of the velocity space, close to the resonance,
which also makes Eqs. (\ref{Norm-1})-(\ref{Norm-4}) easier to solve
numerically. Finally, the spatial dependence is solved for analytically in (%
\ref{Norm-1})-(\ref{Norm-4}), which also simplifies the numerics.

\section{Remarks on the linear theory}

Before we start with a numerical study, let us first make a few comments
regarding the linear theory. As the equation stands, making a linearization
of Eqs. (\ref{Norm-1})-(\ref{Norm-4}) completely removes the quantity $%
\delta v_{q}$ which encodes the quantum properties. Eq. (\ref{Norm-2}) can
then be integrated according to%
\begin{equation}
g_{1}=e^{-ivt}\int_{0}^{t}\hat{\Phi}(t^{\prime })e^{ivt^{\prime }}dt^{\prime
}+e^{-ivt}g_{1}(t=0).  \label{Lin-solution}
\end{equation}%
For certain initial conditions $g_{1}(t=0)$ the integrals in Eqs. (\ref%
{Norm-1}) and (\ref{Lin-solution}) can be solved analytically (see e.g. Ref. 
\cite{Linear-ref} for details), and for these cases we indeed have linear
damping $\hat{\Phi}\propto e^{-t}$ which is written $e^{-\gamma _{L}t\text{ }%
}$in non-normalized units, where 
\begin{equation}
\gamma _{L}=-\frac{\pi }{k}\frac{[\partial G_{0}(v_{z})/\partial
v_{z}]_{\omega /k}}{\int_{\mathrm{nr}}\frac{\partial f_{0}/\partial v_{z}}{%
(\omega -k_{z}v_{z})^{2}}d^{3}v}.  \label{Lin-damping}
\end{equation}%
The reason no effects due to the quantum treatment is seen here is the
assumption of a modest quantum regime $kv_{t}^{2}/\omega \gg \delta v_{%
\mathrm{res}}\gg \hbar k/2m.$ These conditions apply for a resonance in the
tail of the distribution, when the wavelength is not too short. Whenever
these inequalities hold we can also use 
\begin{equation}
G_{0}(v_{z}+\hbar k/2m)-G_{0}(v_{z}-\hbar k/2m)\simeq \frac{\hbar k}{m}\frac{%
\partial G_{0}}{\partial v_{z}}  \label{lin-approx}
\end{equation}
which is the reason the finite difference is replaced by the classical
expression containg a derivative in Eq. (\ref{Lin-damping}). Since $g_{0}$, $%
g_{1},g_{2}...$ varies on a much shorter scale length in velocity space as
compared to $G_{0}$, we note that a similar approximation cannot be applied
for these quantities. We stress that in a general scenario (e.g. for a
beam-plasma system or for the regime $\hbar k^{2}/m\sim \omega $) the
approximation (\ref{lin-approx}) may give an incorrect description of linear
Landau damping, but in our case the linear regime is classical to a good
approximation. It shold also be noted that the evolution of $g_{1}$ exhibits
phase mixing due to the factors $e^{-ivt}$ (see also Fig. 1 of the next
section) whether or not the linear damping expression contains a finite
difference or a derivative. Hence the quantum influence on the regime of
linear Landau damping tend to be relatively modest also when the
approximation (\ref{lin-approx}) is avoided.

\section{Numerical solutions}

\begin{figure}[tbph]
\centering
\includegraphics[width=0.5\textwidth]{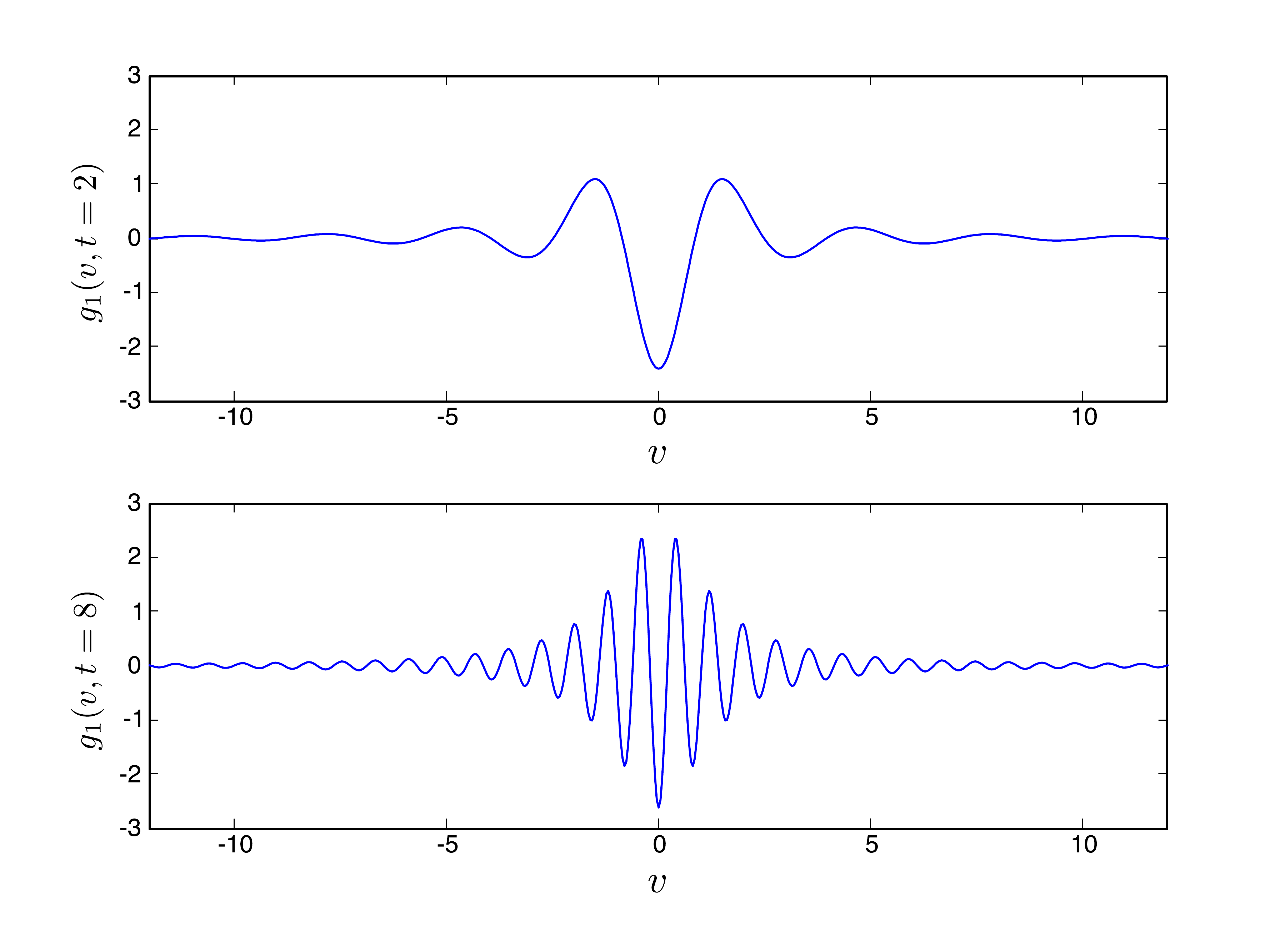}
\caption{The real part of part of $g_{1}$ evaluated in the small amplitude
regime for $t=2$ (upper panel) and $t=8$ (lower panel). The term $ivg_{1}$
of Eq. (\protect\ref{Norm-2}) leads to phase mixing, i.e. the development of
increasingly small scales in velocity space. It should be noted that quantum
effects does not counteract this behavior for small amplitudes.}
\label{fig:wp}
\end{figure}

\begin{figure}[tbph]
\centering
\includegraphics[width=0.5\textwidth]{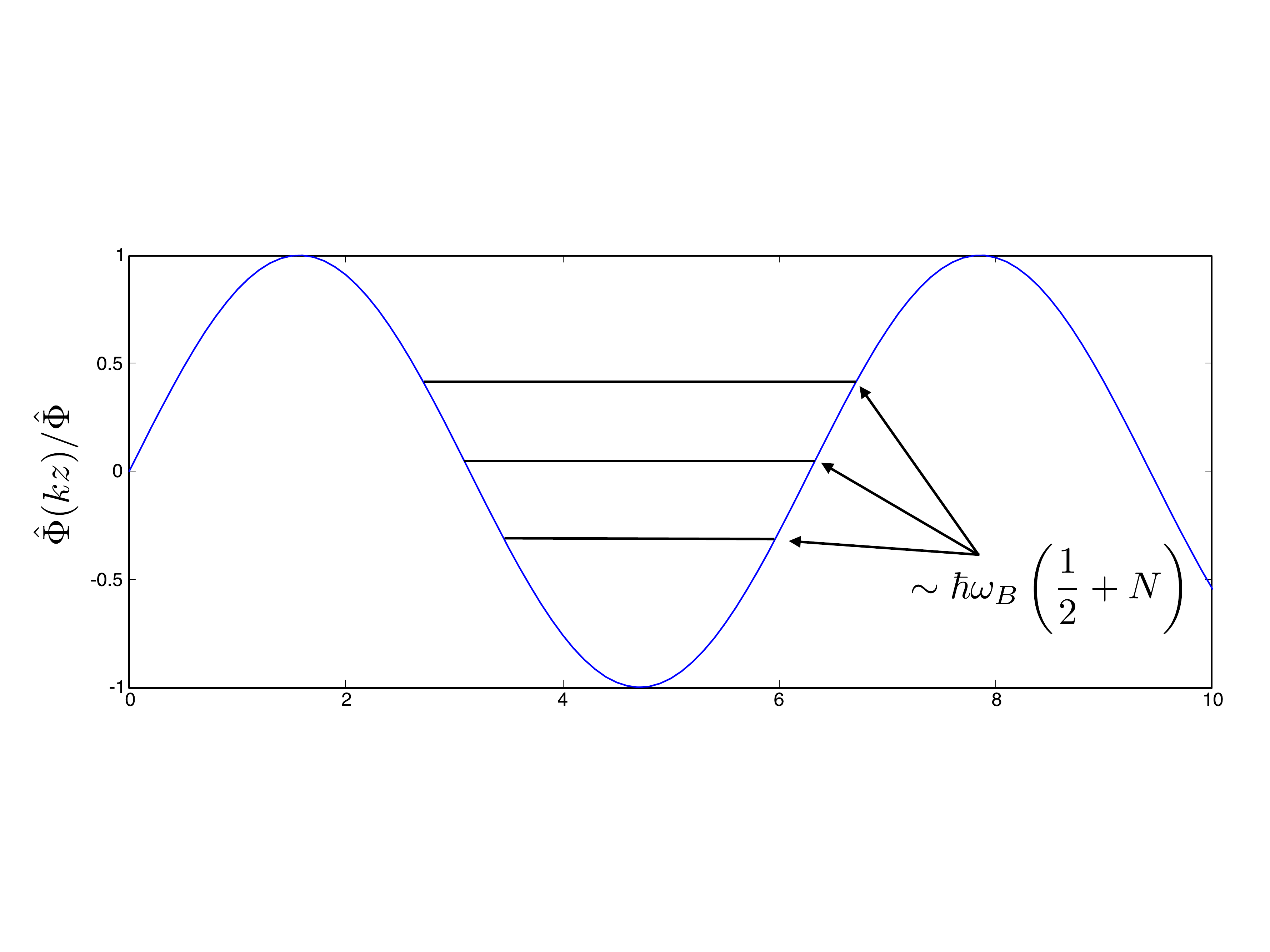}
\caption{A qualitative illustration of the trapped particles. For a fixed
wave amplitude particles trapped particles in the wave field have discrete
energies, with an energy step of the order of $\hbar \protect\omega _{B}$.
When this becomes comparable to $q\hat{\Phi}$ only a few states are trapped,
and quantum mechanical effects become important to describe the evolution.}
\label{fig:wp}
\end{figure}

The system (\ref{Norm-1})-(\ref{Norm-4}) has been studied numerically using
a staggered leapfrog finite differencing technique. Two classes of initial
conditions have been used. In the first case $g_{1}(v,t=0)=\hat{\Phi}%
(t=0)/(iv+1)$ and $g_{0}(t=0)=g_{n}(t=0)$ $=0$ (for $n\geq 2$). This has the
advantage that the Wigner function approaches the value outside the
resonance reegion for $v\gg 1$. The other choice is to put also $%
g_{1}(v,t=0)=0$ such that the initial Wigner function is identically zero in
the resonant region. It turns out that the evolution is almost completely
independent of this difference in initial conditions, if the initial
amplitude $\hat{\Phi}(t=0)$ is increased in the latter case such as to make
the initial energy equal in the two cases. In what follows all numerical
results will refer to the case with $g_{1}(v,t=0)=\hat{\Phi}(t=0)/(iv+1)$.\
Picking a small initial amplitude $\hat{\Phi}(t=0)\ll 1$ the system shows a
damping $\hat{\Phi}\propto e^{-t}$ as expected due to the normalization of
the time variable. As noted in the previous section the evolution is
independent of the quantum parameter $\delta v_{q}$, which is related to the
assumption $\hbar \omega /mv_{t}^{2}\ll 1$ that was made in the derivation.
Fig. 1 compares the real part of $g_{1}$ for different times. The evolution
shows the evolution towards smaller scale lengths in velocity space due to
phase mixing. When nonlinearities come into play this \ process plays part
in increasing the relative importance of quantum effects in wave-particle
interaction, as mathematically the system (\ref{Norm-1})-(\ref{Norm-4})
differs from the classical limit once the scale length of $g_{n}$ in
velocity space is smaller than $\delta v_{q}$. However, a more physical way
to understand why quantum effects become important more easily in the
nonlinear regime than in the linear regime is shown in Fig. 2. Here as a
rough approximation we have described the discrete eigenstates of trapped
particles as that of an harmonic oscillator where the eigenfrequency is the
bounce frequency $\omega _{B}$. Naturally this is rather crude, as the
electrostatic potential is only harmonic for the lowest energy states (if at
all), and moreover the potential can vary dynamically to a smaller or larger
degree depending on the parameters of the problem. Still the simple picture
in Fig 2 is sufficient to identify one of the key parameters $R_{\mathrm{tr}%
} $, which is the ratio of the trapping potential over the energy quanta of
trapped particles, $R_{\mathrm{tr}}=q\hat{\Phi}/\hbar \omega _{B}$. If the
initial values have $R_{\mathrm{tr}}\gg 1$ the evolution of the wave
amplitude $\hat{\Phi}(t)$ is classical to a good approximation. On the other
hand, when $R_{\mathrm{tr}}$ decreases towards unity the discrete energy
states of the trapped particles will modify the evolution significantly. In
terms of our normalized variables we note that the quantum trapping
parameter can be written as $R_{\mathrm{tr}}=q\hat{\Phi}/\hbar \omega
_{B}=m\omega _{B}/\hbar k^{2}=\hat{\Phi}^{1/2}/\delta v_{q}$. For $R_{%
\mathrm{tr}}<1$ we will not have trapped particles. However, as we will see
below the close coupling between the existence of trapped particles and
nonlinear evolution that holds in the classical regime does not generally
apply in the quantum regime.

\begin{figure}[tbph]
\centering
\includegraphics[width=0.5\textwidth]{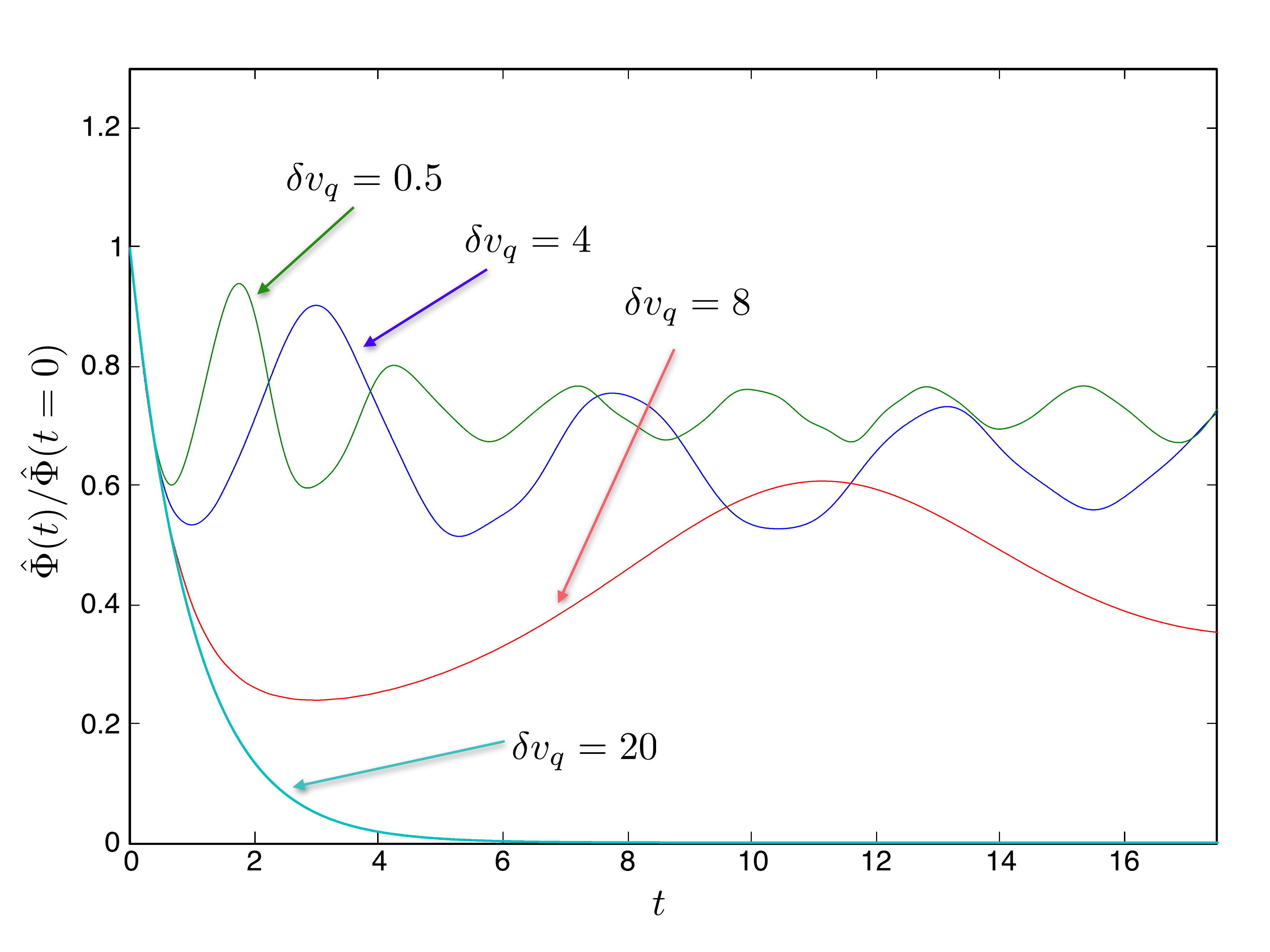}
\caption{The evolution of $\hat{\Phi}(t)$ for initial amplitude $\hat{\Phi}%
(t=0)=4$ and $\protect\delta v_{q}=0.5$, $4$, $8$ and $20$. The value $%
\protect\delta v_{q}=0.5$ gives a classical evolution to a good
approximation, whereas for $\protect\delta v_{q}=20$ the nonlinearities are
almost completely suppressed by the quantum effects.}
\label{fig:wp}
\end{figure}

\begin{figure}[tbph]
\centering
\includegraphics[width=0.5\textwidth]{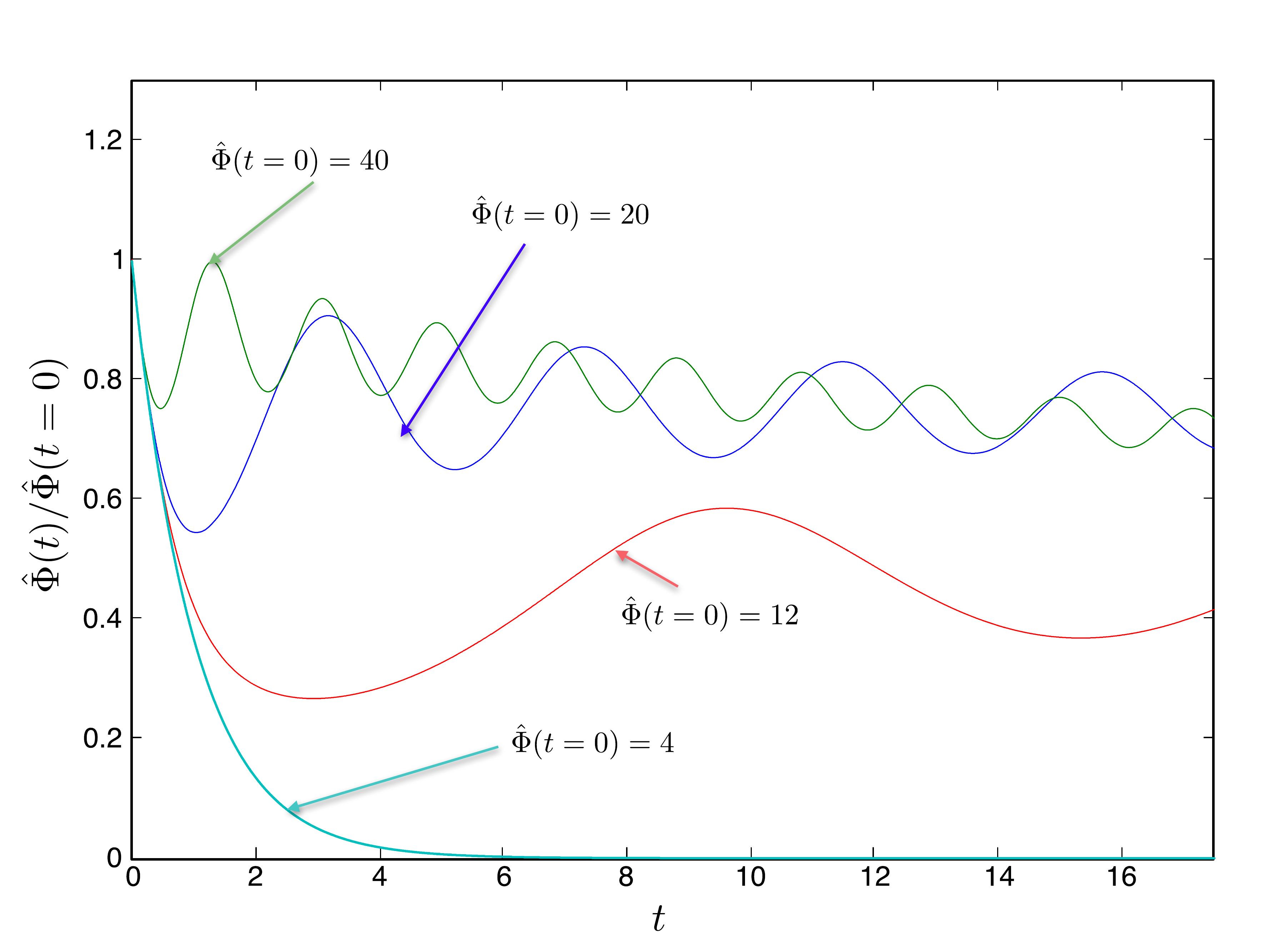}
\caption{ The evolution of $\hat{\Phi}(t)$ for $\protect\delta v_{q}=20$.
The inital amplitudes are $\hat{\Phi}(t=0)=4$, $12$, $20$ and $40$.}
\label{fig:wp}
\end{figure}

Next we investigate how the value of the quantum parameter $\delta v_{q}$
affects the evolution of $\hat{\Phi}(t)$ in the nonlinear regime. Keeping
the normalized initial amplitude equal to $\hat{\Phi}(t=0)=4$ (which means
that the (initial) bounce frequency is $\omega _{B}=2\gamma _{L}$), we
follow the evolution up to $t=17.5$ for various values of $\delta v_{q}$.
The result is shown in Fig. 3. For $\delta v_{q}\lesssim 1$ the evolution of
the wave amplitude more or less coincides with the classical case. This
means that an initial drop in amplitude is followed by oscillations with a
frequency of the order of the bounce frequency $\omega _{B}$. As $\omega
_{B} $ is not a constant, it is no surprise that these oscillations are
slightly irregular. This result is in agreement with previous works on the
classical case, see e.g. \cite{Brodin-1997,ONeil-1965,Manfredi-1997}.\ When
the quantum parameter $\delta v_{q}$ is increased, more of the wave energy
is converted to the particles (the initial amplitude drop is larger) and the
period of the amplitude oscillations become longer. Eventually when $\delta
v_{q}=20$ the initial amplitude drop is so large such that the evolution has
become almost completely linear. Keeping $\delta v_{q}=20$ and instead
varying the initial amplitude, the evolution $\hat{\Phi}(t)$ for various
values of $\hat{\Phi}(t=0)$ is shown in Fig 4. The same qualitative features
as in Fig 3 can be seen. That is the amplitude oscillation gets a lower
frequency with decreasing $\hat{\Phi}(t=0)$, and the initial drop in wave
amplitude is larger, until eventually for small enough initial amplitude the
nonlinearities are suppressed and we obtain linear damping. A more
quantitative analysis based on Figs 3 and 4 reveals that amplitude
oscillations have a frequency $\omega _{\mathrm{amp}}\sim (\omega
_{B}^{2}/(1+\delta v_{q})-\gamma _{L}^{2}/2)^{1/2}$ Classically $\delta
v_{q}=hk^{2}/2m\gamma _{L}$ is small and $\omega _{\mathrm{amp}}\sim \omega
_{B}$ for $\omega _{B}/\gamma _{L}\gg 1$. When the quantum velocity shift $%
\delta v_{q}$becomes comparable to the characteristic scale of $g_{1}$ the
frequency of the amplitude oscillations decreases. Still the amplitude may
oscillate nonlinearly even for large $\delta v_{q}$ and $R_{\mathrm{tr}}\ll
1 $ (in the regime of no trapped particles). The transition to linear
evolution occur when the expression for $\omega _{\mathrm{amp}}$ becomes
begative, in which case the amplitude oscillation frequency decreases too
fast for the nonlinear oscillations to get started.

Next we look closer on some of the details of the numerical results. As
shown in Figs 3 and 4, the evolution of the wave amplitude changes rather
smoothly with $\delta v_{q}$. However, the various harmonics of the Wigner
function $g_{n}$ is more sensitive to the change in $\delta v_{q}$. The
discrete structure involving momentum changes $\hbar k=2m\delta v_{q}$ can
be seen directly in the Wigner function, perhaps most clearly in $g_{0}$. In
a very rough sense there is a relation between the energy states displayed
in Fig. 2 and the momentum structure shown in the upper panel of Fig. 5. As
shown in Fig 5 where $g_{0}(v,t=10)$ is plotted in the classical and quantum
mechanical regime, respectively, there is a very distinct difference between
the quantum mechanical and classical regime. For the former case clear dips
for $v\approx -2\delta v_{q}$ and $v\approx -\delta v_{q}$ can be seen, as
well as peaks for $v\approx \delta v_{q}$ and $v\approx 2\delta v_{q}$. It
should be noted, however, that a discrete structure in the momentum
dependence can be seen even in the absence of trapped particles.

Another important aspect is the convergence in the sum over harmonics $g_{n}$%
, which is much faster in the quantum regime. A comparison of the classical
and quantum regime is displayed in Fig. 6. We see that the relative amount
of energy in the third harmonic $\int \left\vert g_{3}(t)\right\vert
^{2}dv/W_{\mathrm{tot}}$ is changed dramatically when $\delta v_{q}$ is
changed from $0.5$ to $4$. The convergence in the sum over the harmonics is
determined by the trapping parameter $R_{\mathrm{tr}}$. In the classical and
nonlinear regime with $R_{\mathrm{tr}}\gg 1$ we need to include harmonics up
to $g_{10}$, or even more. Besides the value of $R_{\mathrm{tr}}$ the number
of harmonics needed depends on the degree of the nonlinearity and how long
the evolution is followed. In the quantum regime when $R_{\mathrm{tr}}>1$
including up to $g_{3}$ is typically more than enough to get convergence.
This applies even when $\omega _{\mathrm{amp}}\gg \gamma _{L}$ and the
evolution is strongly nonlinear. While the trapping parameter $R_{\mathrm{tr}%
}$ determines much of the properties of the Wigner function in the resonant
region, it is the parameter $R_{\mathrm{nl}}=\omega _{B}^{2}/[\gamma
_{L}^{2}(1+\delta v_{q})]=\hat{\Phi}/(1+\delta v_{q})$ that determines the
degree of nonlinearity. In the nonlinear regime $R_{\mathrm{nl}}\gtrsim 1/2$%
\thinspace\ this parameter determines the frequency of the nonlinear
amplitude oscillations, and we have $\omega _{\mathrm{amp}}^{2}=(R_{\mathrm{%
nl}}-1/2)\gamma _{L}^{2}$. For small $R_{\mathrm{nl}}$ (in practice $R_{%
\mathrm{nl}}<0.25$) the evolution is linear to a very good approximation.
The reason why $R_{\mathrm{nl}}$ determines the nonlinearity can be
understood roughly as follows: A key time-scale is the time-scale $t_{%
\mathrm{char}}$ for modifying the background distribution. Classically this
time-scale is set by the inverse bounce-frequency, i.e. we have $t_{\mathrm{%
char}}^{2}\sim \omega _{B}^{-2}\propto $ $\hat{\Phi}^{-1}$. Mathematically
this simply reflects the fact that the coupling to $g_{0}$ and the higher
harmonics $g_{n}$ is linear in the wave amplitude, as seen in Eqs (\ref%
{Norm-2})-(\ref{Norm-4}). When quantum effects enters the velocity
derivative is replaced by a finite difference. Classically when the scale
lengths in velocity space increases (cf Fig. 1) the magnitude of the
nonlinear terms increases. However, the finite value of $\delta v_{q}$
prevents the continuous increase of the nonlinear coupling, as is seen from
Eqs (\ref{Norm-2})-(\ref{Norm-4}), and effectively the transition from
classical to quantum regime correspond to a change $\hat{\Phi}\rightarrow 
\hat{\Phi}/(1+\delta v_{q})$. Physically this means that the characteristic
time for acceleration of particles increases when the minimum velocity
change comes in steps of $2\delta v_{q}$. \ Noting that $\gamma _{L}^{2}t_{%
\mathrm{char}}^{2}>1$ for the nonlinear modifications of $g_{0}$ to occur
faster than the linear damping explains the significance of the parameter $%
R_{\mathrm{nl}}$.

\begin{figure}[tbph]
\centering
\includegraphics[width=0.5\textwidth]{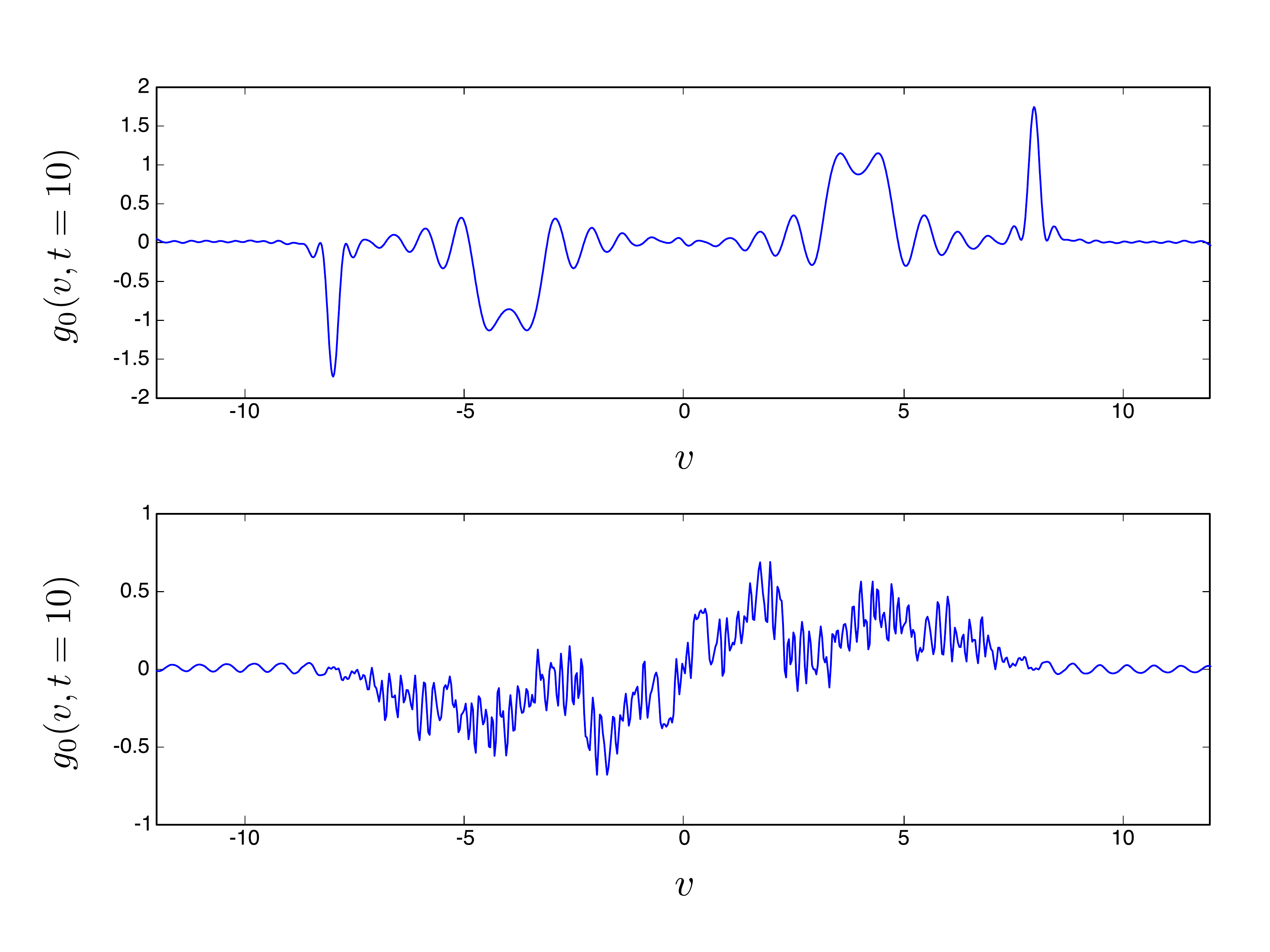}
\caption{The profile of $g_{0}(v,t=10)$ shown for initial amplitude $\hat{%
\Phi}(t=0)=4$ for $\protect\delta v_{q}=0.5$ (lower panel) and for $\protect%
\delta v_{q}=4$ (upper panel). The discrete nature of the veloctiy
dependence with peaks at discret values of $\protect\delta v_q$ is clear in
the quantum regime. There is also a relation to the discrete energy states
of Fig. 2, since the discrete velocity structure disappears when there is a
large number of trapped states, as seen in the lower panel. However, there
is not a one-to-one correspondence between the energy states and the
velocity structure. Specifically the discrete structure of the velocity
dependence can be present in the strong quantum regime when the particle
trapping is absent due to quantum effects.}
\label{fig:wp}
\end{figure}

The results reported here are similar in some respects to the findings of
Ref. \cite{SFR-1991} and Ref. \cite{Daligault-2014}. In Ref. \cite{SFR-1991}
the full Wigner equation was solved numerically. It was then found that the
nonlinear regime of Landau damping was suppressed when $\hbar k/mv_{t}=8$.
This is a rather extreme quantum regime, and our condition for suppressing
the nonlinear regime (essentially $\omega _{\mathrm{amp}}^{2}$ being
negative) which roughly gives $\hbar k^{2}/2m>\omega _{B}^{2}/\gamma _{L}$
is much easier to fulfill. The nmain reason for our relaxed quantum
condition is that we have focused on a resonance in the tail of the
distribution, where $\gamma _{L}\ll \omega $. In this case $\delta v_{q}$
become comparable to the velocity scale length close to the resonance long
before $\delta v_{q}$ becomes comparable to the thermal (or Fermi) velocity.
Thus we can investigate a regime where the wave-particle interaction is
quantum mechanical, although the real part of the wave frequency is
determined by the classical dispersion relation. By contrast, for a
resonance in the bulk of the distribution, all quantum effects appear
simultaneously. Next we compare our results with those of Ref. \cite%
{Daligault-2014}, that has studied the initial evolution of the wave
particle interaction. We note that that the comparison of the quantum
time-scales $t_{q}=2m/\hbar k^{2}$ with the classical bounce time $t_{B}$ $%
=2\pi /\omega _{B}$ made in \cite{Daligault-2014} is very similar to the
discussion made above. In particular Ref. \cite{Daligault-2014} notes that
the condition $t_{q}\ll $ $t_{B}$ implies the absence of trapped particles.
However, since Ref. \cite{Daligault-2014} only studies the initial
evolution, the conclusion that the nonlinear regime is suppressed for $%
t_{q}<t_{B}$ suggested in that work is not accurate. Using the notation of 
\cite{Daligault-2014}, rather the condition for suppressing the nonlinear
regime is $t_{q}<t_{B}^{2}\gamma _{L}$, as explained above.

Much of the above features can be summarized in Fig. 7 that shows the
different regimes of Landau damping plotted as a function of the
nonlinearity parameter $\omega _{B}/\gamma _{L}$ and the quantum parameter $%
\hbar k^{2}/2m\gamma _{L}$. The interesting part of the diagram is the
nonlinear quantum regime $\omega _{B}/\gamma _{L}\geq 1$ and $\hbar
k^{2}/2m\gamma _{L}\geq 1$. It should be noted that the condition for
quantum suppression of the nonlinear amplitude oscillations given by Ref. 
\cite{Daligault-2014} is the same condition that separates our region V from
region VI (i.e. the condition that controls the existence of trapped
particles). Interestingly, however, we find that the system can undergo
nonlinear bounce-like oscillations even without trapped particles, and hence
the region of quantum suppression (region IV) is determined by a distinct
condition involving the quantum modified bounce frequency. It should be
noted that the collisional influence can modify the different regimes to
some extent. This will be discussed in some detail in the final section.

A feature of Fig. 4 not explained by the above discussion is the small but
continuous decrease of wave energy seen for the highest amplitude in Fig. 4.
Besides the amplitude oscillation there is a continuous decrease in $\hat{%
\Phi}(t)$ for initial condition $\hat{\Phi}(t=0)=40$ but not for $\hat{\Phi}%
(t=0)=20$. To understand this we must investigate how the energy of the
particles in the resonance region is distributed among the harmonics. As
described above the convergence is the sum over harmonics is fast for $R_{%
\mathrm{tr}}=\sqrt{\hat{\Phi}}/\delta v_{q}\ll 1$. $\ $For the curves
displayed in Fig 4 only the one with $\hat{\Phi}(t=0)=40$ has significant
harmonic generation. For all the other curves the harmonics $g_{n}$ for $%
n\geq 2$ is effectively not excited, and the two terms $(1/2)\int_{\mathrm{%
res}}g_{0}^{2}dv+$ $\int_{\mathrm{res}}\left\vert g_{1}^{2}\right\vert dv$
accounts for more than $99\%$ of the particle energy. However, due to the
rapid phase mixing of $g_{2}$ for large $\delta v_{q}$, this harmonics does
not return the particle energy to wave energy, but instead the second
harmonic acts as a leakage of energy, not taking part in bounce-like
oscillations. This can be contrasted against the oscillatory evolution of
the energy of the third harmonic for $\delta v_{q}=0.5$ which could be seen
in Fig. 6. To a small extent the continuous loss of wave energy to higher
harmonics can be observed also for $\hat{\Phi}(t=0)=20$, but to a much
smaller degree. This is illustrated in Fig 8$\,$, where the relative energy
content in the second harmonic $\int_{\mathrm{res}}\left\vert
g_{2}^{2}\right\vert dv/W_{\mathrm{tot}}$ is plotted for $\hat{\Phi}(t=0)=20$
and for $\hat{\Phi}(t=0)=40$. As can be seen, by a comparison with Fig. 4
the continuous increase of energy in the second harmonic accounts rather
well for the overall downward trend in wave energy. A similar mechanism in
principle exists also for lower values of $\delta v_{q}$. However here it is
much less effective. The reason is twofold. Firstly, for lower $\delta v_{q}$
the peaks of $g_{2}$ and higher harmonics occurs for a smaller velocity, and
phase mixing is less effective. Thus the energy of the higher harmonics can
be returned back to lower harmonics more easily. Secondly, \ for the same
degree of nonlinearity (same value of $R_{\mathrm{nl}}$) the trapping
parameter is larger for smaller $\delta v_{q}$. Hence a larger number of
harmonics are excited for lower $\delta v_{q}$. This means that the highest
harmonic excited takes a smaller proportion of the energy, in which case the
leakage of energy is smaller. In practice, the details of the energy loss is
likely to be determined by collisional effects, as very small scales in
velocity space develops during the evolution \cite{Brodin-1997}.

\begin{figure}[tbph]
\centering
\includegraphics[width=0.5\textwidth]{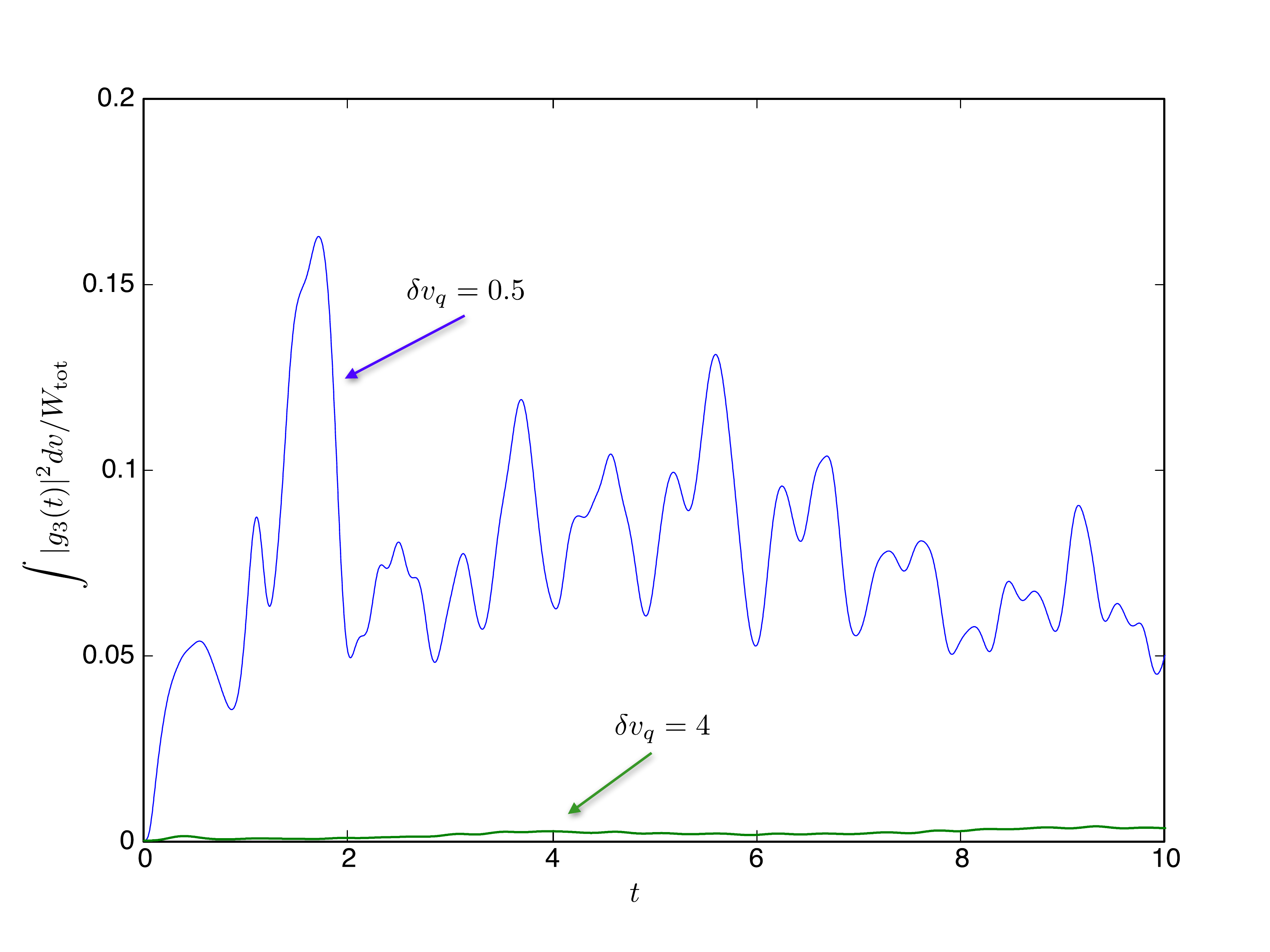}
\caption{Comparison of the relative energy in the third harmonic $\protect%
\int \left\vert g_{3}(t)\right\vert ^{2}dv/W_{\mathrm{tot}}$ for initial
amplitude $\hat{\Phi}(t=0)=4$. The two curves are computed for $\protect%
\delta v_{q}=0.5$ and for $\protect\delta v_{q}=4$ respectively.}
\label{fig:wp}
\end{figure}

\begin{figure}[tbph]
\centering
\includegraphics[width=0.5\textwidth]{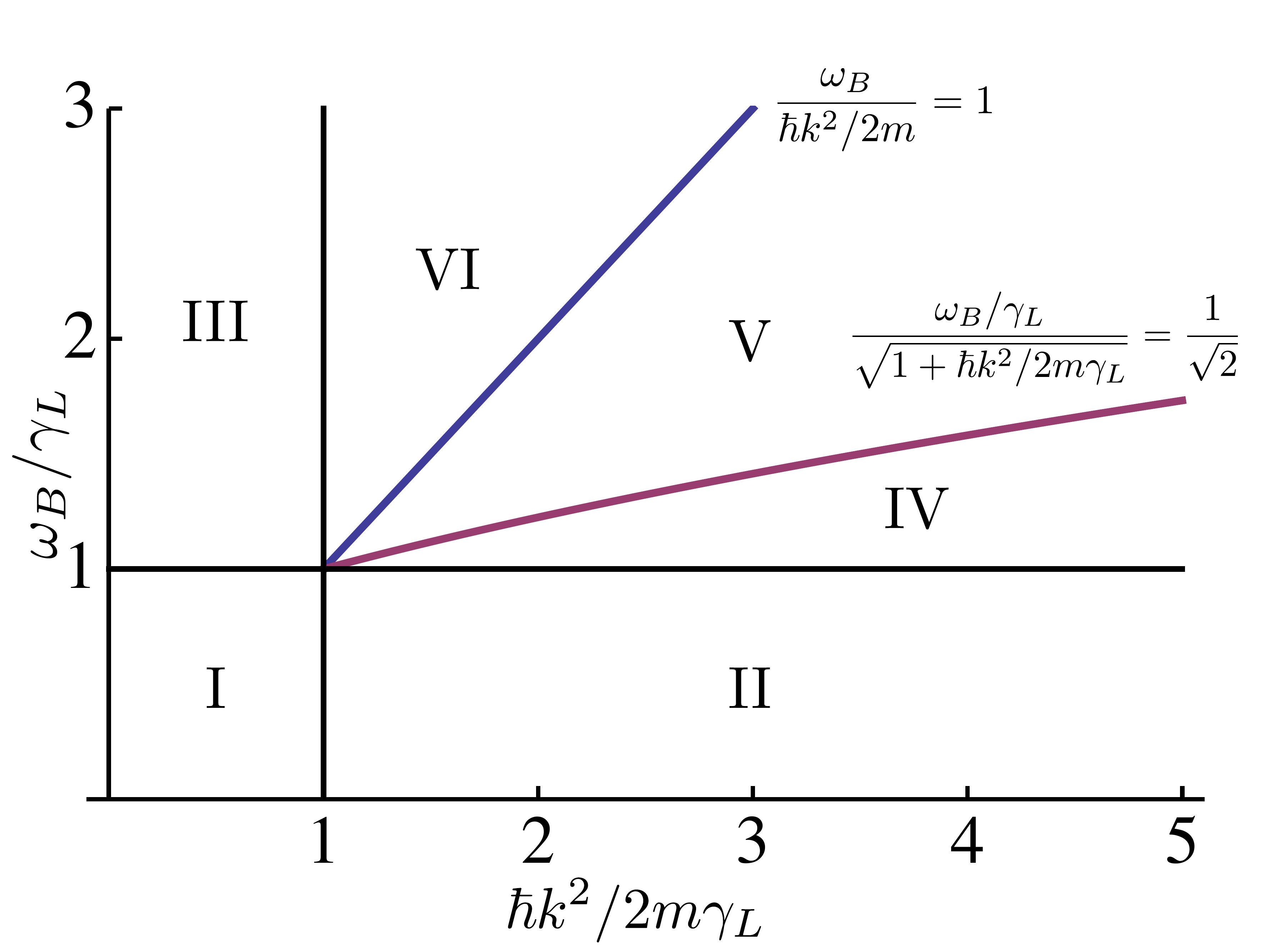}
\caption{Summary of the different regimes of Landau damping for a resonance
in the tail. Region I and II: Linear Landau damping. The quantum effects in
region II does not change the value of the linear damping for a resonance in
the tail Region III: Classical nonlinear regim with bounce oscillation of
frequency $\sim \protect\omega_B$. Region IV: Quantum suppression of
nonlinear bounce-like oscillations. Region V: Nonlinear regime with
bounce-like oscillations in the absence of trapped particles Region VI:
Region of trapped particles with quantum modified bounce oscillations. Note
that the same expression $\sim \protect\omega_B/(1+\hbar k^2/m\protect\gamma%
_L)^{1/2} $ for the quantum modified amplitude oscillations apply in both
region V and VI }
\label{fig:regimes}
\end{figure}

\begin{figure}[tbph]
\centering
\includegraphics[width=0.5\textwidth]{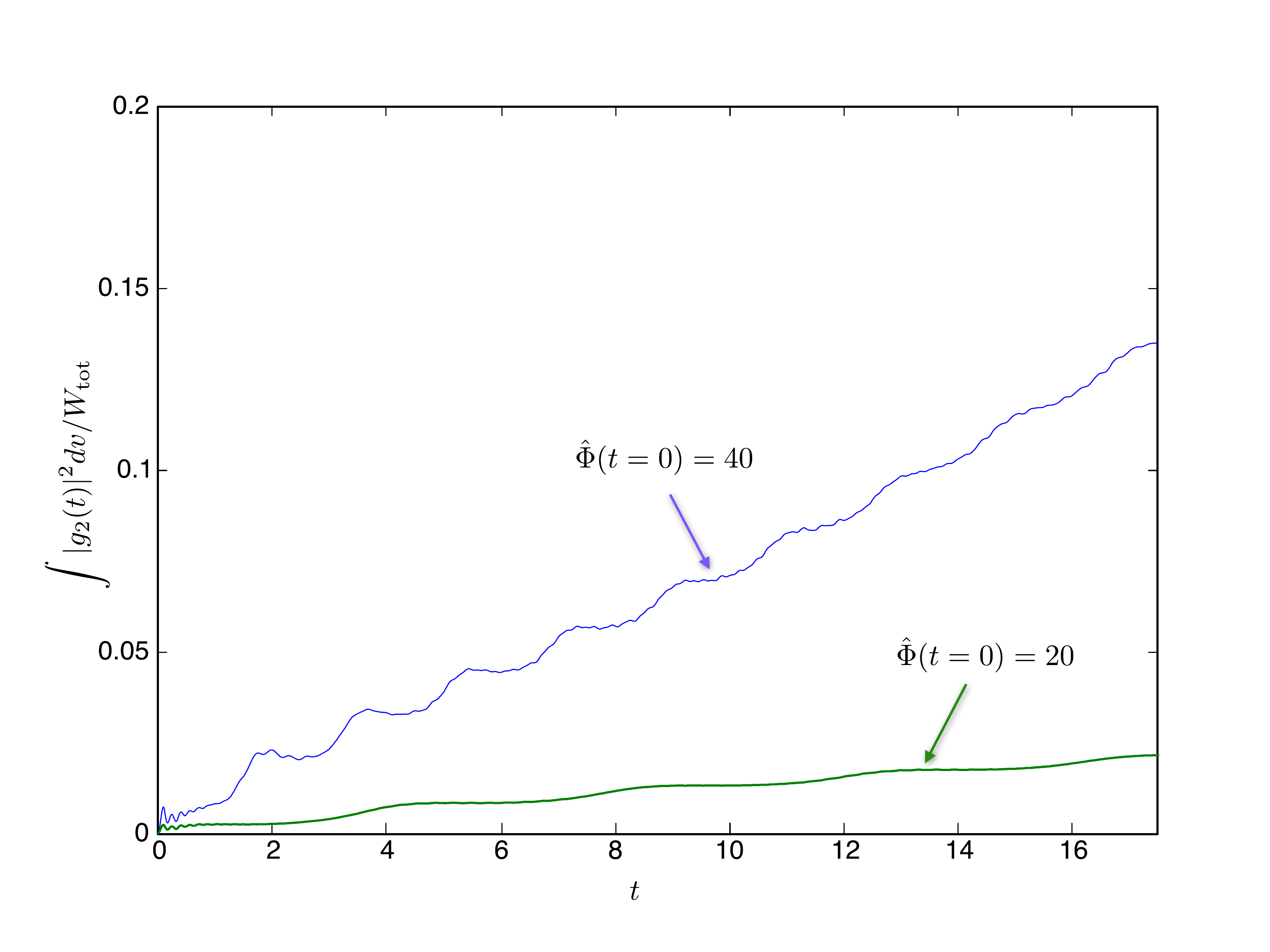}
\caption{Comparison of the relative energy in the second harmonic $\protect%
\int \left\vert g_{2}(t)\right\vert ^{2}dv/W_{\mathrm{tot}}$ for $\protect%
\delta v_{q}=20$. The two curves are computed for initial amplitude $\hat{%
\Phi}(t=0)=20$ and $\hat{\Phi}(t=0)=40$ respectively.}
\label{fig:wp}
\end{figure}

\section{Summary and Discussion}

Usually particle dispersive effects as accounted for by the Wigner-Moyal
equation are assumed to be significant when the thermal de Broglie length $%
\lambda _{\mathrm{dB}}=\hbar /mv_{\mathrm{t}}$, or its counterpart for a
degenerate plasma, $\hbar /mv_{\mathrm{F}}$ (where $v_{\mathrm{F}}$ is the
Fermi velocity),$~$becomes comparable to the Debye length $\lambda _{\mathrm{%
D}}=v_{\mathrm{t}}/\omega _{p}$ (or $v_{\mathrm{F}}/\omega _{p}$). The
physics behind this condition is that the importance of collective effects
require scale lengths not much shorter than the Debye length, and for the
sinus-operator in Eq. (\ref{Wigner-2}) not to reduce to its classical limit
we need $\hbar L_{\mathrm{char}}^{-1}/mv_{\mathrm{char}}$ $\sim 1$. If the
characteristic spatial scale length $L_{\mathrm{char}}$ is assumed not to be
smaller than $\lambda _{\mathrm{D}}$, and the velocity scale length $v_{%
\mathrm{char}}$ is of the order of $v_{\mathrm{t}}$ we get the condition $%
\hbar \omega _{p}/mv_{\mathrm{t}}^{2}\sim 1$ (or $\hbar \omega _{p}/mv_{%
\mathrm{F}}^{2}$ $\sim 1$ for a degenerate plasma) for quantum effects to be
significant \cite{manfredi2006,Haas-book,Shukla-Eliasson-RMP}. This
condition applies broadly to many situations, but does not hold for
wave-particle interaction in general, as we have $v_{\mathrm{char}}\ll v_{%
\mathrm{t}}$ in case the resonance lies in the tail of the distribution. In
the present paper, focusing on the case of Langmuir waves of a single
wavelength, we have found that quantum effects become important in the
nonlinear regime once $\hbar k^{2}/m\sim \omega _{B}$ in which case the
trapped particle energies are quantized. Already when $\hbar k^{2}/m>\gamma
_{L}$ nonlinear oscillations at the bounce frequency is slowed down in
accordance with $\omega _{B}\rightarrow \omega _{B}/(1+\delta v_{q})^{1/2}$.
While these oscillations are qualitatively similar to the well-known bounce
oscillations \cite{ONeil-1965}, it should be noted that in the quantum
regime such oscillations can occur even in the absence of trapped particles.
Increasing the quantum parameter $\delta v_{q}$ \ even further, eventually
the nonlinear oscillations are suppressed when $\omega _{B}/(1+\delta
v_{q})^{1/2}<\gamma _{L}$ in which case usual linear damping takes place.

Since the phenomena of study evolves on a time scale slower than the plasma
frequency, the above picture can be modified when the collisional influence
is accounted for. Let us illustrate this by considering a concrete example.
First we note that the long term evolution will always be affected by
collisions. The characteristic time scale for collitions to be important $%
\tau _{\mathrm{c}}$ is given by $\min [\nu _{\mathrm{ei}}^{-1},\nu _{\mathrm{%
ee}}^{-1}]$. \ To some extent we can assure that the phenomena of study
occur on a faster scale by picking a relatively large amplitude and avoding
a resonance too far out in the tail of the distribution. However, for a
strongly coupled plasma this will not suffice, as $\nu _{\mathrm{ee}}$
approaches the plasma frequency rather than being a small parameter. Thus
for our model to apply we must first of all consider a plasma that is weakly
coupled. As an example we pick a plasma with a number density $n_{0}\simeq $ 
$10^{28}\mathrm{m}^{-3}$ and a temperature $T\simeq 10^{7}\mathrm{K}$ that
correspond to a coupling parameter $\Gamma \simeq 0.01$. Such parameter
values may result from laser-plasma experiments, see e.g. Ref. \cite%
{glenzer-redmer}. Furthermore the temperature is well above the Fermi
temperature, and we have a Debye length of the order $\lambda _{D}\simeq
10^{-9}\mathrm{m.}$ Adjusting the wavenumber we can chose the amount of
resonant particles that determine the linear Landau damping rate and thereby
pick he quantum parameter. In order to get $\delta v_{q}\simeq 4$ we can aim
at $\gamma _{L}\simeq 10^{12}\mathrm{s}^{-1}$ which is obtained for a wave
number $k\simeq 2\times 10^{8}\mathrm{m}^{-1}$. As $\delta v_{q}>1$ we are
well in the quantum regime. We note, however, that although the plasma is
weakly coupled, the time-scale for collisional damping is still shorter than
the the linear damping time $\gamma _{L}^{-1}$. This is typically the case
in the quantum regime, and hence the long-time evolution is generally much
affected by collisions. However, most of our findings concern the\textit{\
nonlinear} quantum behavior. As we will demonstrate below these findings are
to a large extent still relevant even when collisions are taken into
account. Firstly, the quantum modified bounce frequency $\omega _{B}$ $%
\rightarrow \omega _{B}/(1+\delta v_{q})^{1/2}$ can be experimentally
verified after a few bounce oscillations. For example, if we pick $\hat{\Phi}%
\gtrsim 10\mathrm{V}$ the bounce oscillations take place on a much faster
scale than collisional damping, in which case the quantum modification of
the bounce frequency becomes detectable. Moreover, generally the prediction
of a nonlinear regime in the absence of trapped particles (regime V in Fig.
7) can be verified after a few (quantum modified) bounce oscillations. Hence
in an experimental situation we do not need to follow the evolution until
collisional effects become important. The situation is somewhat different
when it comes to the condition for quantum suppression of the nonlinear
oscillations. It is possible to verify that previously given conditions are
incorrect in agreement with our theory, as the presence of bounce
oscillations at a faster scale than the linear damping rate will confirm
this. However, the quantitative confirmation of our condition that separates
regions IV and V in Fig 7 is likely not possible, as that would require us
to study the system when the quantum modified bounce frequency is of the
same order as the linear Landau damping rate. In this case we need to follow
the evolution long enough such that collisions will modify the picture. Thus
we conclude that most features of Fig 7 remains even when we account for
collisional effects, but that there are restrictions arising from the
collisional influence that makes the boundary between region IV and V
uncertain.

While the study here has focused on Landau damping due to electrons, it
should be noted that a similar process may take place for photons \cite%
{Mendonca-2001,Mendonca-2006}. Photon Landau damping can be relevant to
Langmuir waves with relativistic phase velocities, in the presence of
electromagnetic radiation, when resonant electrons are nearly absent. The
wave nature of the photons provides the classical counterpart to the quantum
electron states, and we can similarly identify a photon bounce frequency $%
\omega _{Bph}\sim \omega _{B}$. Photon trapping effects were experimentally
observed by \cite{Murphy-2006}.

\textbf{Acknowledgement} This paper is dedicated to our good friend and
colleague Professor Lennart Stenflo, celebrating his 75:th birthday.


\begin{thebibliography}{99}
\bibitem{Numerical note} The staggered leapfrog method also works well when
the finite differences in the right hand side is replaced by a velocity
derivative, which corresponds to the classical limit. In the quantum regime
Eqs. (\ref{Norm-2})-(\ref{Norm-4}) just constitute a set of ordinary
differential equations (although it is a large set, since the velocity
resolution must be sufficient), and a fourth order Runge-Kutta method could
also be applied.
\end{thebibliography}
\end{document}